\documentclass{article}
\usepackage{spconf,amsmath,graphicx}
\usepackage{algorithm}
\usepackage{algpseudocode}
\usepackage{booktabs, siunitx} 
\usepackage{hyperref}
\usepackage[capitalize,noabbrev]{cleveref}

\title{HiFi++: a Unified Framework for Bandwidth Extension and Speech Enhancement}
\name{ Pavel Andreev$^{*123}$,  Aibek Alanov$^{*124}$, Oleg Ivanov$^{*1}$, Dmitry Vetrov$^{24}$}
\address{
  $^*$ Equal contribution \ \ \ \ \ 
  $^1$ Samsung Research \\ $^2$ Higher School of Economics, Moscow  \ \ \ \ \ 
  $^3$ Skolkovo Institute of Science and Technology, Moscow  \\
  $^4$ Artificial Intelligence Research Institute, Moscow} 
  
\begin{document}
%
\maketitle

 \newcommand{\mlcell}[2][p{2cm}c]{%
     \begin{tabular}[#1]{@{}c@{}}#2\end{tabular}
 }
 \newcolumntype{H}{>{\setbox0=\hbox\bgroup}c<{\egroup}@{}}

\begin{abstract}
Generative adversarial networks have recently demonstrated outstanding performance in neural vocoding outperforming best autoregressive and flow-based models. In this paper, we show that this success can be extended to other tasks of conditional audio generation. In particular, building upon HiFi vocoders, we propose a novel HiFi++ general framework for bandwidth extension and speech enhancement. We show that with the improved generator architecture, HiFi++ performs better or comparably with the state-of-the-art in these tasks while spending significantly less computational resources. The effectiveness of our approach is validated through a series of extensive experiments.
\end{abstract}
\begin{keywords}
speech enhancement, bandwidth extension
\end{keywords}
\section{Introduction}
\label{intro}

The problem of conditional speech generation has great practical importance.
The applications of conditional speech generation include neural vocoding, bandwidth extension (BWE), speech enhancement (SE, also referred to as speech denoising), and many others.
One recent success in the field of conditional speech generation is related to the application of generative adversarial networks \cite{kumar2019melgan, kong2020hifi}. 
Particularly, it was demonstrated that GAN-based vocoders could drastically outperform all publicly available neural vocoders in both quality of generated speech and inference speed. 
In this work, we adapt the HiFi model \cite{kong2020hifi} to the bandwidth extension and speech enhancement tasks by designing new generator.

The key contribution of this work is a novel HiFi++ generator architecture that allows to efficiently adapt the HiFi-GAN framework to the BWE and SE problems. The proposed architecture is based on the HiFi generator with new modules.
Namely, we introduce spectral preprocessing (SpectralUnet), convolutional encoder-decoder network (WaveUNet) and learnable spectral masking (SpectralMaskNet) to the generator's architecture. 
Equipped with these modifications, our generator can be successfully applied to the bandwidth extension and speech enhancement problems. 
As we demonstrate through a series of extensive experiments, our model performs on par with state-of-the-art 
in bandwidth extension and speech enhancement tasks.
The model is  significantly more lightweight than the examined counterparts while having better or comparable quality.


\section{Background}

\textbf{Bandwidth extension} \ \  Frequency bandwidth extension \cite{liu2022voicefixer, lin2021two} (also known as audio super-resolution) can be viewed as a realistic increase of signal sampling frequency. 
Speech bandwidth or sampling rate may be truncated due to poor recording devices or transmission channels. 
Therefore super-resolution models are of significant practical relevance for telecommunication.

For the given audio $ x = \{x_i \}_{i = 1}^{N} $ with the low sampling rate  $s$, a bandwidth extension model aims at restoring the recording in high resolution $ y = \{x_i \}_{i = 1}^{N \cdot S / s} $ with the sampling rate $ S $  (i.e., expand the effective frequency bandwidth). 
We generate training and evaluation data by applying low-pass filters to a high sample rate signal and then downsampling the signal to the sampling rate $s$:
\begin{equation}
x = \mathrm{Resample} (\mathrm{lowpass}(y, s / 2), s, S), 
\end{equation}
 where $ \mathrm {lowpass} (\cdot, s / 2) $ means applying a low-pass filter with the cutoff frequency $ s / 2 $ (Nyquist frequency at the sampling rate $ s $), $ \mathrm{Resample} (\cdot, S, s) $ denotes downsampling the signal from the sampling frequency $ S $ to the frequency $s$. Following recent works \cite{wang2021towards, liu2022voicefixer}, we randomize low-pass filter type and order during training for model robustness. 


\noindent \textbf{Speech enhancement} \ \   Audio denoising \cite{fu2021metricgan+, tagliasacchi2020seanet} is always a major interest in audio processing community because of its importance and difficulty. 
In this task, it is required to clean the original signal (most often speech) from extraneous distortions. 
We use additive external noise as distortion. 
Formally speaking, given the noisy signal $x = y + n$ the denoising algorithm predicts the clean signal $y$, i.e. suppresses the noise $n$.

\begin{figure*}[!ht]
\begin{center}

\includegraphics[width=1.0\linewidth]{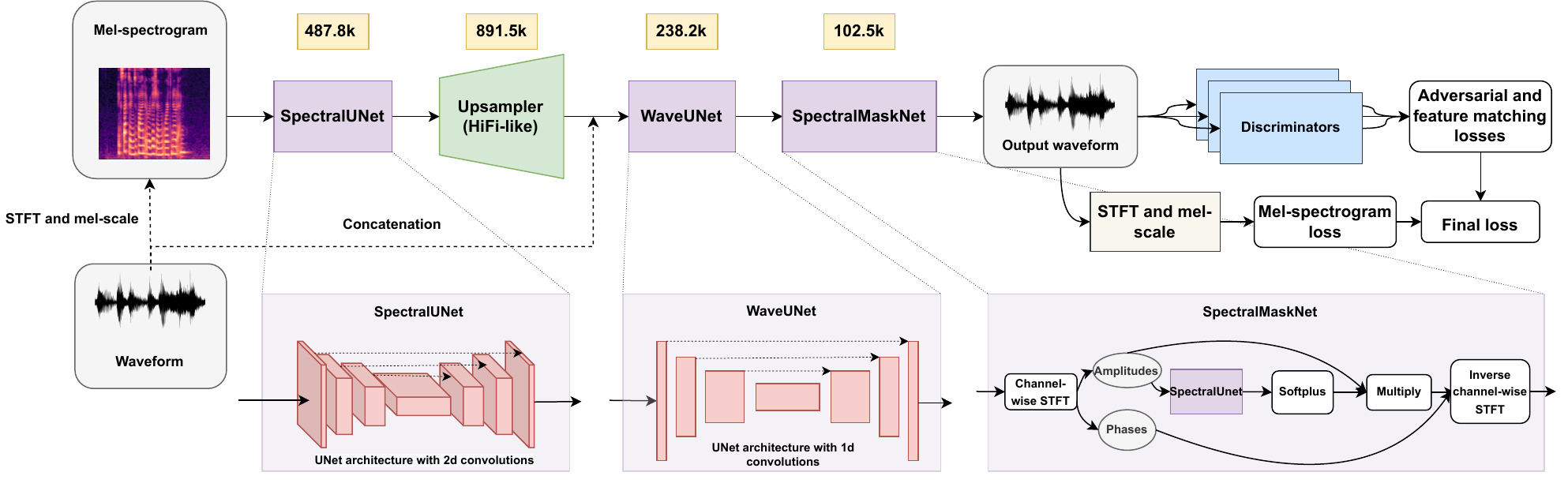}
\end{center}

\caption{HiFi++ architecture and training pipeline. 
The HiFi++ generator consists of the HiFi-like Upsampler and three introduced modules SpectralUNet, WaveUNet and SpectralMaskNet (their sizes are in yellow boxes). 
The generator's architecture is identical for BWE and SE. 
}
\label{fig:hifi_plus}
\end{figure*}

\section{HiFi++}

\subsection{Adapting HiFi-GAN Generator For Bandwidth Extension and Speech Enhancement}



In this paper, we propose a novel HiFi++ architecture that adapts HiFi generator \cite{kong2020hifi} to the SE and BWE problems by introducing new modules: SpectralUNet, WaveUNet and SpectralMaskNet (see \Cref{fig:hifi_plus}). 
The HiFi++ generator is based on the HiFi part ($V2$ version of HiFi-GAN generator) that takes as an input the enriched mel-spectrogram representation by the SpectralUNet and its output goes through postprocessing modules: WaveUNet corrects the output waveform in time domain while SpectralMaskNet cleans up it in frequency domain. We also tried to change the order of WaveUNet and SpectralMaskNet modules and did not observe significant improvements.  
We describe the introduced modules in details in the next paragraphs.

 \noindent  \textbf{SpectralUNet} \ \  We introduce the SpectralUNet module as the initial part of the HiFi++ generator that takes the input mel-spectogram (see \Cref{fig:hifi_plus}).
The mel-spectrogram has a two-dimensional structure and the two-dimensional convolutional blocks of the SpectralUnet model are designed to facilitate the work with this structure at the initial stage of converting the mel-spectrogram into a waveform.
The idea is to simplify the task for the remaining part of the HiFi++ generator that should transform this 2d representation to the 1d sequence. 
We design the SpectralUNet module as UNet-like architecture with 2d convolutions. This module also can be considered as the preprocess part that prepares the input mel-spectrogram by correcting and extracting from it the essential information that is required for the desired task. 

\noindent  \textbf{WaveUNet} \ \ The WaveUNet module is placed after the HiFi part (Upsampler) and takes several 1d sequences concatenated with the input waveform as an input. This module operates directly on time domain and it can be considered as a time domain postprocessing mechanism that improves the output of the Upsampler and merges the predicted waveform with the source one. The WaveUNet module is an instance of the well-known architecture Wave-U-Net \cite{stoller2018wave} which is a fully convolutional 1D-UNet-like neural network. 
This module outputs the 2d tensor which consists of $m$ 1d sequences that will be processed and merged to the output waveform by the next SpectralMaskNet module.

 \noindent  \textbf{SpectralMaskNet} \ \  We introduce the SpectralMaskNet as the final part of the generator which is a learnable spectral masking. It takes as an input the 2d tensor of $m$ 1d sequences and applies channel-wise short-time Fourier transform (STFT) to this 2d tensor. Further, the SpectralUNet-like network takes the amplitudes of the STFT output (in the case of vocoding it takes also the output of SpectralUNet module concatenated) to predict multiplicative factors for these amplitudes. The concluding part consists of the inverse STFT of the modified spectrum (see \Cref{fig:hifi_plus}). Importantly, this process does not change phases. The purpose of this module is to perform frequency-domain postprocessing of the signal. 
We hypothesize that it is an efficient mechanism to remove artifacts and noise in frequency domain from the output waveform in a learnable way. Note that similar techniques have been used in speech enhancement literature as a standalone solution \cite{wisdom2019differentiable}.



\subsection{Training objective}


We use the multi-discriminator adversarial training framework that is based on \cite{kong2020hifi} for time-domain models' training. However, instead of multi-period and multi-scale discriminators  we use several identical discriminators that are based on multi-scale discriminators but operate on the same resolutions and have smaller number of weights (we reduce channel number in each convolutional layer by the factor of 4).
We employ three losses, namely LS-GAN loss $\mathcal{L}_{GAN}$ \cite{mao2017least}, feature matching loss $\mathcal{L}_{FM}$ \cite{ kumar2019melgan}, and mel-spectrogram loss $\mathcal{L}_{Mel}$ \cite{kong2020hifi}:
\begin{align}
    \mathcal{L}(\theta) &= \mathcal{L}_{GAN}(\theta) + \lambda_{fm}\mathcal{L}_{FM}(\theta) + \lambda_{mel}\mathcal{L}_{Mel}(\theta) \\
    \mathcal{L}(\varphi_i) &= \mathcal{L}_{GAN}(\varphi_i), \quad i = 1, \dots, k.
\end{align}
where $\mathcal{L}(\theta)$ denotes loss for generator with parameters $\theta$, $\mathcal{L}(\varphi_i)$ denotes loss for i-th discriminator with parameters $\varphi_i$ (all discriminators are identical, except initialized differently). In all experiments we set $\lambda_{fm} = 2$, $\lambda_{mel} = 45$, $k=3$. 

\section{Experiments}
\label{exps}

All training hyper-parameters and implementation details will be released with source codes. 


\subsection{Data}


 \noindent  \textbf{Bandwidth extension} \ \  We use publicly available dataset VCTK \cite{yamagishi2019cstr} (CC BY 4.0 license) which includes 44200 speech recordings belonging to 110 speakers. We exclude 6 speakers from the training set and 8 recordings from the utterances corresponding to each speaker to avoid text level and speaker-level data leakage to the training set. For evaluation, we use 48 utterances corresponding to 6 speakers excluded from the training data. Importantly, the text corresponding to evaluation utterances is not read in any recordings constituting training data. 

 \noindent  \textbf{Speech denoising} \ \ We use VCTK-DEMAND dataset \cite{valentini2017noisy} (CC BY 4.0 license) for our denoising experiments.  The train sets (11572 utterances) consists of 28
speakers with 4 signal-to-noise ratio (SNR) (15, 10, 5, and 0
dB). The test set (824 utterances) consists of 2 speakers with 4 SNR (17.5, 12.5, 7.5, and 2.5 dB). Further details can be found in the original paper.

\subsection{Evaluation}
\label{sec:evaluation}

\noindent  \textbf{Objective evaluation} \ \  We use conventional metrics WB-PESQ \cite{rix2001perceptual}, STOI \cite{taal2011algorithm}, scale-invariant signal-to-distortion ratio (SI-SDR) \cite{le2019sdr}, DNSMOS \cite{reddy2022dnsmos} for objective evaluation of samples in the SE task. In addition to conventional speech quality metrics,
we considered absolute objective speech quality measure based
on direct MOS score prediction by a fine-tuned wave2vec2.0
model (WV-MOS), which was found to have better system-level
correlation with subjective quality measures than the other objective metrics\footnote{\href{https://github.com/AndreevP/wvmos}{https://github.com/AndreevP/wvmos}}.



 \noindent  \textbf{Subjective evaluation} \ \  We employ 5-scale MOS tests for subjective quality assessment. All audio clips were
normalized to prevent the influence of audio volume differences on the raters. 
The referees were restricted to be english speakers with proper listening equipment. 

\begin{table*}[!h]
\centering
\caption{Bandwidth extension results on VCTK dataset. * indicates re-implementation.}
	\label{table:bwe}
	\scalebox{1.0}{
  \begin{tabular}{ lllHH|llHH|llHH|lH }
    \toprule
        & \multicolumn{4}{c}{\textbf{BWE (1kHz)}} & \multicolumn{4}{c}{\textbf{BWE (2kHz) }} & \multicolumn{4}{c}{\textbf{BWE (4kHz) }} && \\
    \midrule
    
   Model & MOS & \mlcell{WV-\\MOS} & STOI & PESQ & MOS & \mlcell{WV-\\MOS} & STOI & PESQ& MOS & \mlcell{WV-\\MOS} & STOI & PESQ & \mlcell{\# Param \\ (M)} & \mlcell{\# MACs \\ (G)} \\
    \midrule\midrule
Ground truth               &      $4.62 \pm 0.06$   &  4.17  &  1.00 &  4.64 &       $4.63 \pm 0.03$  &  4.17 &   1.00 &   4.64 &          $4.50 \pm 0.04$  & 4.17  &   1.00 &   4.64 &    - & -\\
\midrule
HiFi++ (ours) &    $\mathbf{4.10 \pm 0.05}$      &  \textbf{3.71} &  0.86 &  1.74 &           $\mathbf{4.44 \pm 0.02}$ & \textbf{3.95 } &   0.94 &   2.54 &   $\mathbf{4.51  \pm  0.02}$ &        4.16 &      1.00 &   3.74 &    \textbf{1.7} &  \textbf{2.8} \\
SEANet  &    $3.94  \pm  0.09$ &      3.66 &   0.82 &  1.54 &    $\mathbf{4.43 \pm 0.05}$ &       \textbf{3.95} &      0.93 &   2.43 &     $ 4.45 \pm  0.04 $ &     \textbf{4.17} &    0.99 &   3.65 &    9.2 &  4.5 \\
VoiceFixer      &      $3.04  \pm  0.08$ &     3.21 &  0.73 &  1.44 &         $3.82 \pm 0.06$ & 3.50 &     0.78 &   1.73 &    $ 4.34 \pm  0.03 $&      3.77 &   0.83 &   2.38 &  122.1 & 34.4 \\
TFiLM          &       $1.98 \pm 0.02$&     1.65 &  0.81 &  2.11 &           $2.67 \pm 0.04$ &  2.27 &   0.91 &   2.63 &       $ 3.54 \pm  0.04 $ &      3.49 &  1.00 &   3.52 &   68.2 & - \\
\midrule
input            &       $1.87 \pm 0.08$ &   0.39 &    0.78 &  2.60 &       $2.46 \pm 0.04$ &     1.74 &  0.88 &   3.04 &    $ 3.36 \pm  0.06 $ &       3.17 &    0.99 &   3.65 &    - &  -\\

    \bottomrule
  \end{tabular}
  }
  \vspace{-0.3cm}
\end{table*}

\subsection{Bandwidth Extension}

In our bandwidth extension experiments, we use recordings with a sampling rate of 16 kHz as targets and consider three frequency bandwidths for input data: 1 kHz, 2kHz, and 4 kHz. The models are trained independently for each bandwidth.
The results and comparison with other techniques are outlined in \cref{table:bwe}. Our model HiFi++ provides a better tradeoff between model size and quality of bandwidth extension than other techniques. Specifically, our model is 5 times smaller than the closest baseline SEANet \cite{li2021real} while outperforming it for all input frequency bandwidths. In order to validate the superiority of HiFi++ over SEANet in addition to MOS tests we conducted pair-wise comparisons between these two models and observe statistically significant dominance of our model (p-values are equal to $2.8 \cdot 10^{-22}$ for 1 kHz bandwidth, $0.003$ for 2 kHz, and $0.02$ for 4 kHz for the binomial test). 

Importantly, these results highlight the importance of adversarial objectives for speech frequency bandwidth extension models. Surprisingly, the SEANet model \cite{li2021real} appeared to be the strongest baseline among examined counterparts leaving the others far behind. This model uses adversarial objective similar to ours. The TFilm \cite{birnbaum2019temporal} and 2S-BWE \cite{lin2021two} models use supervised reconstruction objectives and achieve very poor performance, especially for low input frequency bandwidths.

\subsection{Speech Enhancement}

The comparison of the HiFi++ with baselines is demonstrated in the \cref{table:se}. Our model achieves comparable performance with state-of-the-art models VoiceFixer \cite{liu2022voicefixer} and DB-AIAT \cite{yu2022dual} counterparts while being dramatically more computationally efficient. Interestingly, VoiceFixer achieves high subjective quality while being inferior to other models according to objective metrics, especially to SI-SDR and STOI. Indeed, VoiceFixer doesn't use waveform information directly and takes as input only mel-spectrogram, thus, it misses parts of the input signal and is not aiming at reconstructing the original signal precisely leading to poor performance in terms of classic relative metrics such as SI-SDR, STOI, and PESQ. Our model provides decent relative quality metrics as it explicitly uses raw signal waveform as model inputs. At the same time, our model takes into account signal spectrum, which is very informative in speech enhancement as was illustrated by the success of classical spectral-based methods. It is noteworthy that we significantly outperform the SEANet \cite{tagliasacchi2020seanet} model, which is trained in a similar adversarial manner and has a larger number of parameters, but does not take into account spectral information.

An interesting observation is the performance of the MetriGAN+ model \cite{fu2021metricgan+}. While this model is explicitly trained to optimize PESQ and achieves high values of this metric, this success does not spread on other objective and subjective metrics.

\begin{table*}[!ht]
\centering
\caption{Speech denoising results on Voicebank-DEMAND dataset. * indicates re-implementation.}
	\label{table:se}
  \begin{tabular}{ lllHHHH|ll }
    \toprule
   Model & MOS & WV-MOS & SI-SDR & STOI & PESQ & DNSMOS  & \mlcell{\# Par \\ (M)} & \mlcell{\# MACs \\ (G)} \\
    \midrule\midrule
 Ground truth &  $4.46 \pm 0.05$  &         4.50 &  - &  1.00 &  4.64 & 3.15 &             - & -\\
 \midrule
 DB-AIAT  & $\mathbf{4.40 \pm 0.05}$ & \textbf{4.38} & \textbf{19.4} & \textbf{0.96} & \textbf{3.27}  & \textbf{3.18} & 2.8 & 41.8  \\
HiFi++ (ours) & $4.31 \pm 0.05$&        4.36 &   17.9 &  0.95 &  2.90 & 3.10 &       \textbf{1.7} &  \textbf{2.8}\\
 VoiceFixer       &  $4.21 \pm 0.06$  &     4.14 &  -18.5 &  0.89 &  2.38 & 3.13 &    122.1 & 34.4 \\
DEMUCS &     $4.17 \pm 0.06$   &  4.37 &   18.5 &  0.95 &  3.03 & 3.14 &      60.8 &  38.1 \\

MetricGAN+      &  $3.98 
\pm 0.06$  &     3.90 &    8.5 &  0.93 &	3.13 & 2.95 &       2.7 &  28.5\\
\midrule
Input            &  $ 3.45 \pm 0.07$ &      2.99 &    8.4 &  0.92 &	1.97	 & 2.53     &      - &  -\\

    \bottomrule
  \end{tabular}
   \vspace{-0.3cm}
\end{table*}


\subsection{Ablation Study}

To validate the effectiveness of the proposed modifications, we performed the ablation study of the introduced modules SpectralUNet, WaveUNet and SpectralMaskNet. For each module, we consider the architecture without this module \textit{with increased capacity} of HiFi generator part to match the size of the initial HiFi++ architecture. 

The results of the ablation study are shown in \cref{table:abl}, which reveal how each module contributes to the HiFi++ performance. We also compare against vanilla HiFi generator model which takes mel-spectrogram as the only input. The structure of the vanilla HiFi generator is the same as in $V1$ and $V2$ versions from HiFi-GAN paper, except the parameter "upsample initial channel" is set to 256 (it is 128 for $V2$ and 512 for $V1$). We can see that WaveUNet and SpectralMaskNet are essential components of the architecture, as their absence notably degrades the model performance. SpectralUNet has no effect on quality of SE and minor positive effect on BWE (statistical significance of improvement is ensured by pairwise test). However, since we match the number of parameters for ablation models with HiFi++, this positive effect comes at no cost, thus, it is useful to include SpectralUNet into generator architecture.

\begin{table}[!h]
\centering
\caption{Ablation study results.}
	\label{table:abl}
	\scalebox{1.0}{
  \begin{tabular}{ llHHH|lHHHHHHHHH }
    \toprule
        & \multicolumn{4}{c}{\textbf{BWE (1kHz)}} & \multicolumn{4}{c}{\textbf{SE }} & &&& && \\
    \midrule
    
   Model & MOS & \mlcell{WV-\\MOS} & STOI & PESQ & MOS & \mlcell{WV-\\MOS} & STOI & PESQ& MOS & \mlcell{WV-\\MOS} & STOI & PESQ & \mlcell{\# Param \\ (M)} & \mlcell{\# MACs \\ (G)} \\
    \midrule\midrule
Ground truth               &      $4.50 \pm 0.06$   &  4.17  &  1.00 &  4.64 &       $4.48 \pm 0.05$  &  4.50 &   1.00 &   4.64 &          $4.50 \pm 0.04$  & 4.17  &   1.00 &   4.64 &    - & -\\
\midrule
Baseline (HiFi++) &    $\mathbf{3.92  \pm  0.04}$      &  \textbf{3.71} &  0.86 &  1.74 &           $\mathbf{4.27  \pm  0.04}$ & \textbf{4.31 } &   0.94 &   2.54 &   $\mathbf{4.51  \pm  0.02}$ &        4.16 &      1.00 &   3.74 &    1.71 &  \textbf{1.5} \\
w/o SpectralUNet   &    $3.83  \pm  0.06$ &      3.64 &   0.82 &  1.54 &    $\mathbf{4.26  \pm 0.05}$ &       4.29 &      0.93 &   2.43 &     $ 4.45 \pm  0.04 $ &     4.29 &    0.99 &   3.65 &    1.72 &  4.5 \\
w/o WaveUNet      &      $3.46  \pm  0.06$ &     3.55 &  0.73 &  1.44 &         $4.19  \pm  0.03$ & 4.23 &     0.78 &   1.73 &    $ 4.34 \pm  0.03 $&      3.77 &   0.83 &   2.38 &  1.75 & 34.4 \\
w/o SpectralMaskNet &     $3.51  \pm  0.06$ &   3.54 &      0.83 &  1.84 &         $4.17  \pm  0.05$&   4.30 &    0.94 &   2.63 &  $ 4.10 \pm  0.04 $ & 3.96   &           1.00 &   3.26 &    1.74 & 2.2\\
vanilla HiFi   &  $ 3.42  \pm  0.05 $ &    3.43 &   0.85 &  1.82 &        $4.17 \pm  0.04$&    4.23 &    0.94 &   2.36 &    $ 4.27 \pm  0.05 $ &         4.05 &  1.00 &   3.84 & 3.56 & 4.2 \\
\midrule
input            &       $1.69  \pm  0.05$ &   0.39 &    0.78 &  2.60 &       $3.51  \pm  0.06$ &     2.99 &  0.88 &   3.04 &    $ 3.36 \pm  0.06 $ &       3.17 &    0.99 &   3.65 &    - &  -\\

    \bottomrule
  \end{tabular}
  }

\end{table}

\section{Conclusion}


In this work, we introduce the universal HiFi++ framework for bandwidth extension and speech enhancement. We show through a series of extensive experiments that our model achieves results on par with the state-of-the-art baselines on BWE and SE tasks. Remarkably, our model obtains such results being much more effiecient (in some cases by two orders of magnitude) than existing counterparts.

\bibliographystyle{IEEEbib}
\bibliography{strings,refs}

\newpage
\onecolumn
\appendix

\section{Architecture details}
\label{app:arch}

\subsection{Upsampler}
\begin{figure}[h]
\begin{center}
\includegraphics[width=0.7\linewidth]{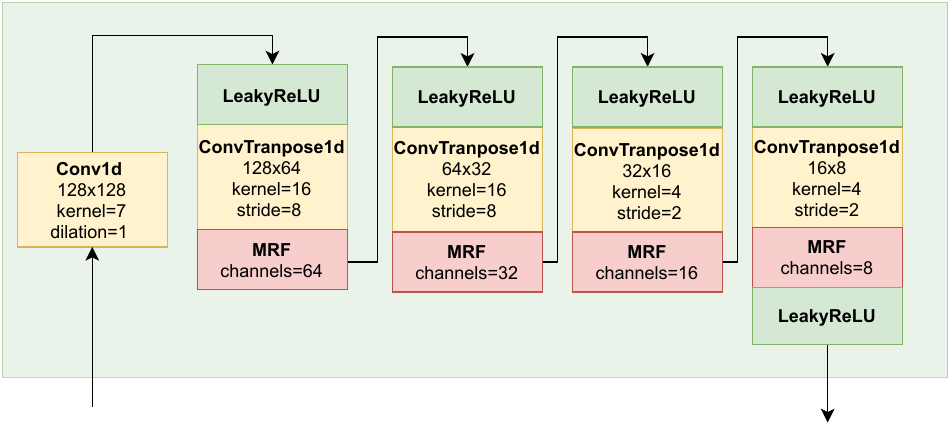}
\end{center}
\vspace{-0.5cm}
\caption{Upsampler architecture.}
\label{fig:arch_hifi}
\end{figure}

\begin{figure}[h]
\begin{center}
\includegraphics[width=0.7\linewidth]{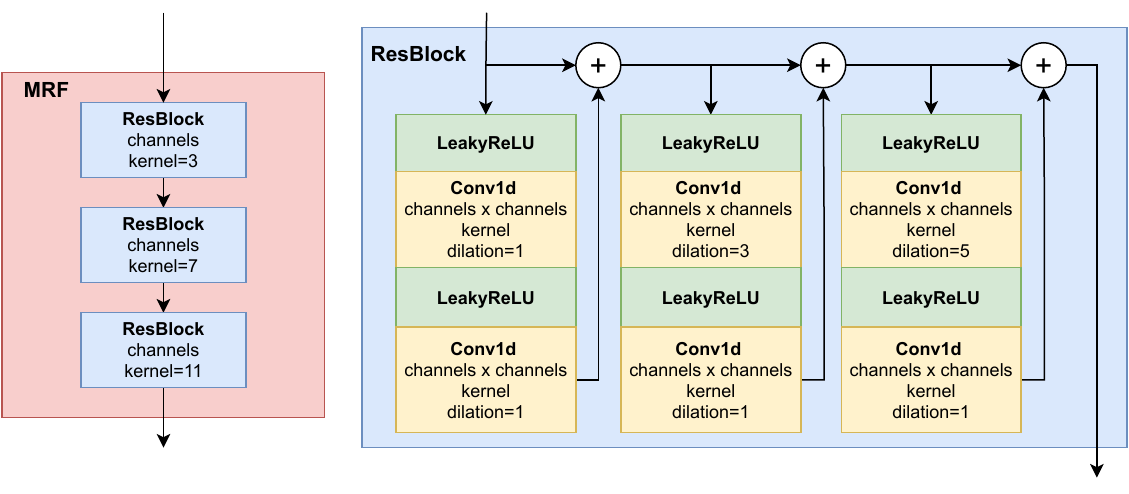}
\end{center}
\vspace{-0.5cm}
\caption{HiFi-GAN generator blocks.}
\label{fig:arch_mrf}
\end{figure}

We use $V2$ configuration of the original HiFi-GAN generator for the mel-spectrogram upsampling part of our model. We outline its architecture on the Figures \ref{fig:arch_hifi} and \ref{fig:arch_mrf}.

\subsection{WaveUNet and SpectralUNet}

\begin{figure}[h!]
    \centering
    \begin{minipage}{0.45\linewidth}
      \centering
      \includegraphics[width=\textwidth]{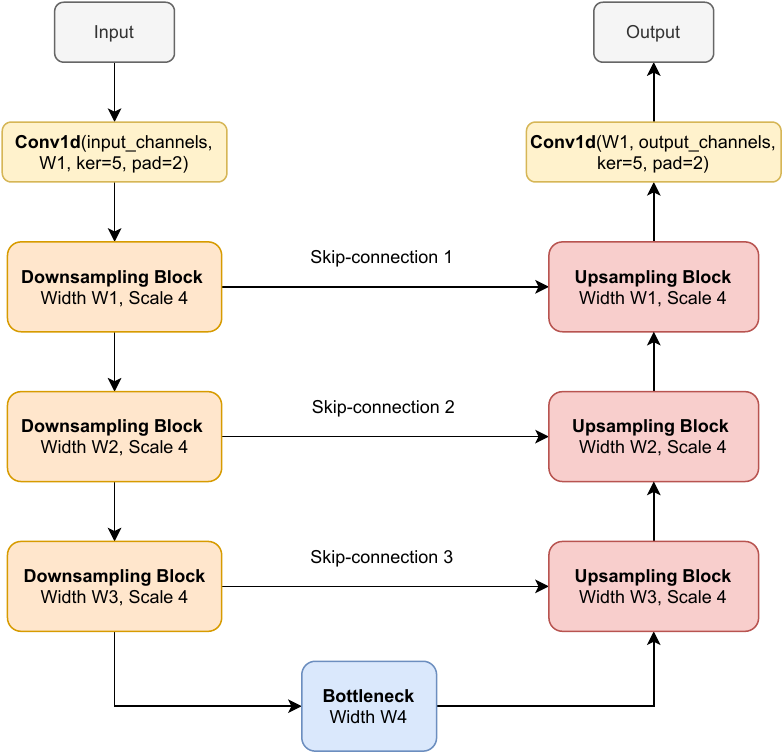}
        \caption{The architecture of WaveUNet model. Block widths [$W1$, $W2$, $W3$, $W4$] are equal to [10, 20, 40, 80].}
\label{fig:arch_waveunet}
    \end{minipage}
    \hfill
    \begin{minipage}{0.45\linewidth}
      \centering
      \includegraphics[width=\textwidth]{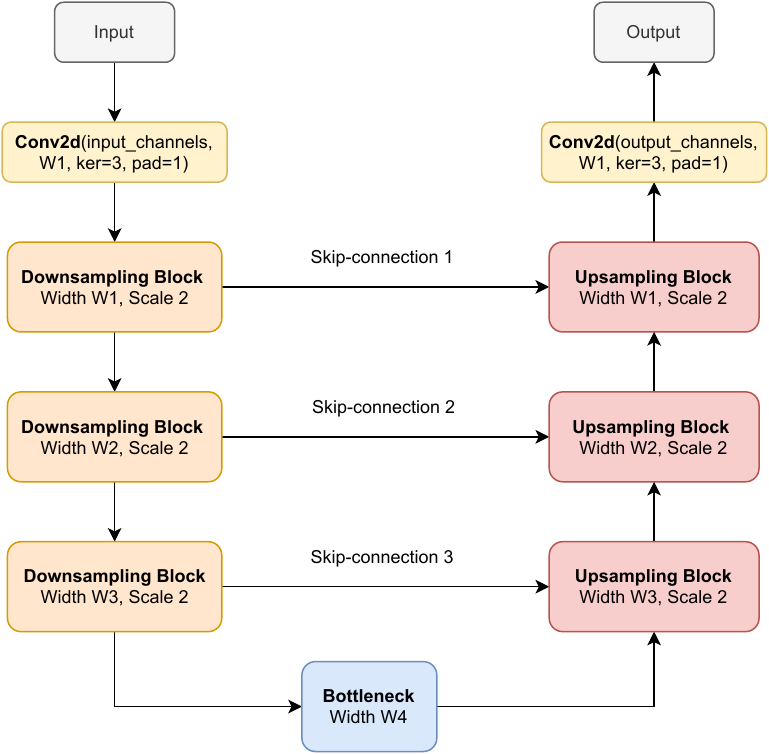}
      \caption{The architecture of SpectralUNet model. Block widths [$W1$, $W2$, $W3$, $W4$] are equal to [8, 16, 32, 64], [8, 12, 24, 32] for SpectralUNet and SpectralMaskNet modules, respectively. }
      \label{fig:arch_specunet}
    \end{minipage}
\end{figure}

\begin{figure}[h]
\begin{center}
\includegraphics[width=1.0\linewidth]{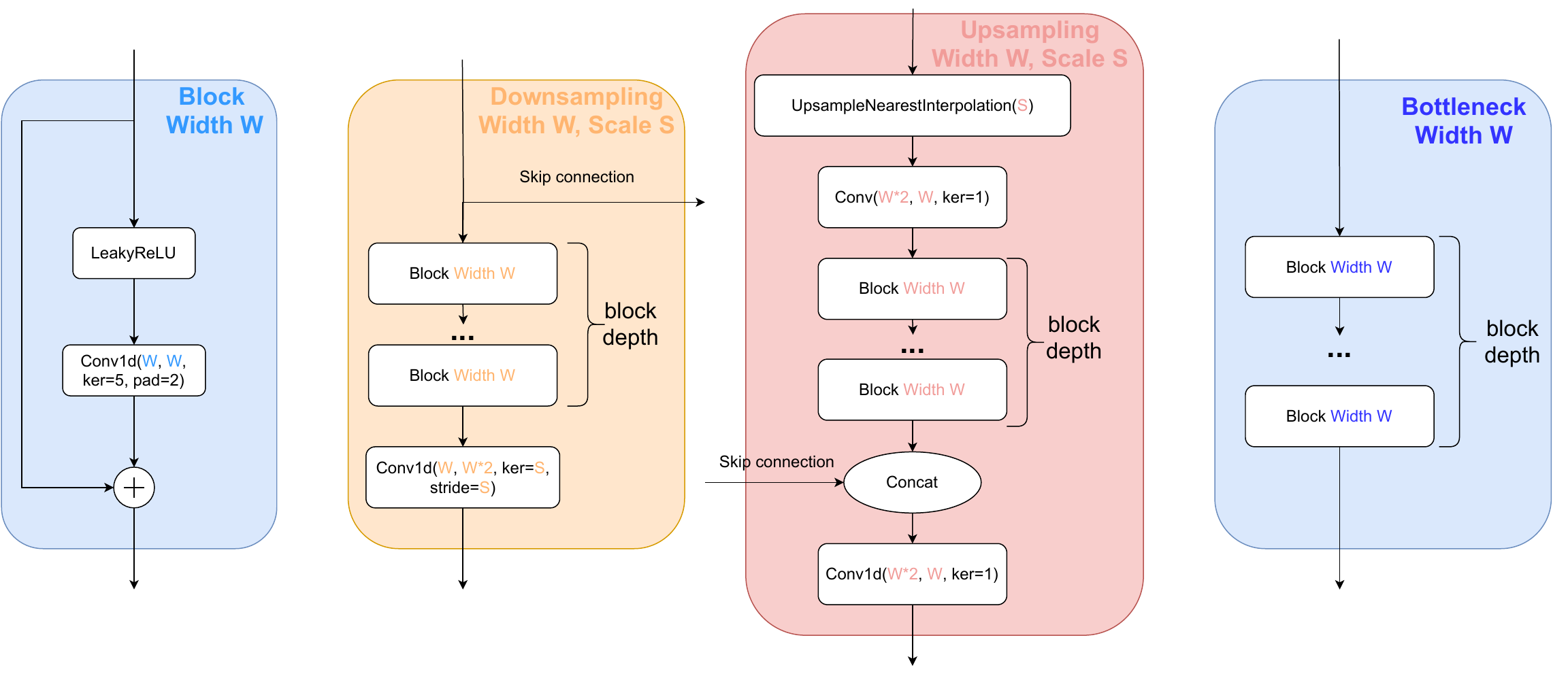}
\end{center}
\caption{WaveUNet blocks. WaveUNet and SpectralUNet blocks share the same architectural structure except SpectralUNet uses 2d convolutions with kernel size $3\times3$ instead of 1d convolutions with kernel size being equal to 5. Block depth is equal to 4.}
\label{fig:arch_unet}
\end{figure}

We use standard fully-convolutional multiscale encoder-decoder architecture for WaveUNet and SpectralUNet networks. The architectures of these networks are depicted on Figures \ref{fig:arch_waveunet} and \ref{fig:arch_specunet}. Each downsampling block of WaveUNet model performes $\times4$ downsampling of the signal across time dimension. Analogously, each downsampling block in SpectralUNet downscales signal $\times2$ across time and frequency dimensions. The width $[W1, W2, W3, W4]$ and block depth parameters control number of parameters and computational complexity of resulting networks. The structure of WaveUNet blocks is outlined on the Figure \ref{fig:arch_unet}.

\section{WV-MOS}
\label{app:wv-mos}
\subsection{MOSNet}
The quality assessment of generated speech is an important problem in the speech processing domain. 
The popular objective metrics such as PESQ, STOI, and SI-SDR are known to be poorly correlated with subjective quality assessment. 
Meanwhile, mean opinion score (MOS) — an audio
quality metric that is measured from human annotators’ feedback is a de-facto standard for speech synthesis systems benchmarking \cite{kong2020hifi} where no objective metrics are usually reported. 
Not only is obtaining MOS expensive, but also it leads to incomparable (from paper to paper) results due to different sets of annotators participating in evaluations.

One recent attempt to address this problem was performed by \cite{lo2019mosnet}.
The paper proposes a deep learning-based objective assessment to model human perception in terms of MOS,
referred to as MOSNet. 
MOSNet used raw magnitude spectrogram as the input feature and three neural network-based models, namely CNN, BLSTM, and CNN-BLSTM are used to extract valuable features from the input and fully connected (FC) layers and pooling mechanisms to generate predicted MOS. 
For training, the authors use MSE Loss with MOS evaluations of VCC 2018 as the targets. 
Among considered architectures of feature extractors, a combination of CNN with BLSTM turned out to be the most preferable on most of the test metrics. 
The advantages of MOSNet over conventional objective speech quality metrics (e.g. PESQ, STOI) are twofold. 
Firstly, it doesn't require a reference signal for quality assessment and predicts MOS score directly from the generated audio. 
Second, deep learning models have great potential in the prediction of subjective quality measurements as it was earnestly shown in the computer vision domain \cite{zhang2018unreasonable}.

\subsection{MOSNet is not able to identify obvious corruptions}

Nevertheless, we found MOSNet model to be a low proficiency speech quality predictor. A simple yet demonstrative example is that MOSNet fails to identify records corrupted by low-pass filters. We applied low pass filters of different cutoff frequencies (1 kHz, 2 kHz, and 4 kHz) to 50 samples from the VCTK dataset and measured the ratio of cases where MOSNet assigned a higher score to the corrupted sample than to the reference one. We also examined additive noise as corruption and measured the analogous failure rate for the Voicebank-Demand test part. The results can be found in \cref{table:mosnet}.

\subsection{WV-MOS: a new absolute objective speech quality measure}

We hypothesized that the MOSNet failure can be to a significant extent attributed to the outdated neural network architecture and lack of pretraining. For this reason, we propose to utilize for MOS prediction a modern neural network architecture wav2vec2.0 \cite{baevski2020wav2vec}. Wav2vec2.0 model is pretrained in a contrastive self-supervised manner, thus, its representations are task-agnostic and can be useful for a variety of downstream tasks. 

Following \cite{lo2019mosnet}, we use listening evaluation results released by Voice Conversion Challenge 2018 \cite{lorenzo2018voice} for model training. We augment pretrained wav2vec2.0 model with a 2-layer MLP head and train the resulting neural network to predict assigned MOS scores. We use mean squared error as training objective, the batch size is equal to 64, the learning rate was set to 0.0001. As a result, we observe higher Spearman’s rank correlation factors on VCC 2018 test set (0.62/0.93 for our model versus 0.59/0.90 for MOSNet on utterance/system levels, respectively). More importantly, we observe our model to better generalize to the audio samples beyond VCC 2018 (see \cref{table:mosnet,table:mosnet_mos,table:mosnet_mos2}).  

\begin{table}[!h]
\centering
\caption{MOS prediction assessment results (toy example). The MOSNet fails to identify such obvious corruption as the application of 4kHz low-pass filters.}
	\label{table:mosnet}
  \scalebox{1.0}{
  \begin{tabular}{ lll }
    \toprule
   Data &  \mlcell{Error rate (\%) \\ (MOSNet)} & \mlcell{Error rate (\%) \\ (WV-MOS)} \\
    \midrule\midrule
    Low-pass filter (1 kHz) & 0 & 0 \\
    Low-pass filter (2 kHz) & 2.1 & 0 \\
    Low-pass filter (4 kHz) & 34.0 & 0 \\
    Voicebank-demand & 12.3 & 3.6 \\
    \bottomrule
  \end{tabular}
  }

\end{table}

We use crowd-sourced studies conducted during this work for validation of the proposed model. Based on crowd-sourced MOS scores we computed utterance-level (i.e., assigned MOSes are averaged for each sample) and system-level (i.e., assigned MOSes are averaged for each model) correlations of the predicted MOS scores with the ones assigned by referees. The results are demonstrated in the \cref{table:mosnet_mos,table:mosnet_mos2}. Interestingly, at the utterance level, the WV-MOS score is correlated much better with the assigned MOS-es for bandwidth extension task while being inferior in this sense to PESQ for vocoding and speech enhancement tasks. At the system level, the situation is more unambiguous as WV-MOS outperforms all examined metrics in terms of Spearman's correlation with assigned MOS scores.

Additionally, we have compared against popular objective metric DNSMOS \cite{reddy2021dnsmos} for speech enhancement. We found that it provides the best system level Spearman's correlation with MOS score in this task (0.72), and competitive utterance-level correlation of 0.61.

\begin{table}[!h]
\centering
\caption{MOS prediction assessment results (crowd-sourced studies). The table shows Spearman’s rank utterance-level correlations with crowd-sourced studies. }

\label{table:mosnet_mos}
\begin{tabular}{ llllllHHHHH}
\toprule
Data & WV-MOS & MOSNet & PESQ &  STOI & SI-SDR & WV-MOS & MOSNet & PESQ &  STOI  & SI-SDR \\
\midrule\midrule
Vocoding & 0.47 & 0.03 & \textbf{0.73} & 0.41 & 0.29 & \textbf{0.96} & 0.11 & 0.89 & 0.45  & 0.57  \\
BWE (1 kHz) & \textbf{0.86} & 0.52 & -0.07 & 0.38  & -0.02 & \textbf{0.97} & 0.67 & -0.17 & 0.55 & -0.07\\
BWE  (2 kHz) & \textbf{0.85} & 0.36 & 0.11 & 0.50 & 0.01 & \textbf{0.90} & 0.71 & -0.02 & 0.57 & -0.12\\
BWE (4 kHz) & \textbf{0.69} & 0.25 & 0.08 & 0.24 & -0.16 & \textbf{0.86} & 0.40 & 0.38 & 0.29 & -0.38\\
SE & 0.64 & 0.44 & \textbf{0.67} & 0.42  & 0.45 & \textbf{0.68} & 0.61 & 0.5 & 0.32 & 0.43\\
\bottomrule
\end{tabular}

\end{table}

\begin{table}[!h]
\centering
\caption{MOS prediction assessment results (crowd-sourced studies). The table shows Spearman’s rank system-level correlations with crowd-sourced studies.}
\label{table:mosnet_mos2}
\begin{tabular}{ lHHHHHlllll}
\toprule
Data & WV-MOS & MOSNet & PESQ &  STOI & SI-SDR & WV-MOS & MOSNet & PESQ &  STOI  & SI-SDR \\
\midrule\midrule
Vocoding & 0.47 & 0.03 & \textbf{0.73} & 0.41 & 0.29 & \textbf{0.96} & 0.11 & 0.89 & 0.45  & 0.57  \\
BWE (1 kHz) & \textbf{0.86} & 0.52 & -0.07 & 0.38  & -0.02 & \textbf{0.97} & 0.67 & -0.17 & 0.55 & -0.07\\
BWE  (2 kHz) & \textbf{0.85} & 0.36 & 0.11 & 0.50 & 0.01 & \textbf{0.90} & 0.71 & -0.02 & 0.57 & -0.12\\
BWE (4 kHz) & \textbf{0.69} & 0.25 & 0.08 & 0.24 & -0.16 & \textbf{0.86} & 0.40 & 0.38 & 0.29 & -0.38\\
SE & 0.64 & 0.44 & \textbf{0.67} & 0.42  & 0.45 & \textbf{0.68} & 0.61 & 0.5 & 0.32 & 0.43\\
\bottomrule
\end{tabular}

\end{table}

\section{Subjective evaluation}
\label{app:subj}

We measure mean opinion score (MOS) of the model using a crowd-sourcing adaptation of the standard absolute category rating procedure.
Our MOS computing procedure is as follows.
\begin{enumerate}
    \item Select a subset of 40 random samples from the test set (once per problem, i. e. for bandwidth extension or speech enhancement).
    \item Select a set of models to be evaluated; inference their predictions on the selected subset.
    \item Randomly mix the predictions and split them into the pages of size 20 almost uniformly.
          Almost uniformly means that on each page there are at least $\lfloor \frac{20}{\mathrm{num\_models}} \rfloor$ samples from each model.
    \item Insert additional 4 trapping samples into random locations on each page: 2 samples from groundtruth, and 2 samples of a noise without any speech.
    \item Upload the pages to the crowd-sourcing platform, set the number of assessors for each page to at least 30.\\
          Assessors are asked to work in headphones in a quiet environment; they must listen the audio until the end before assess it.
    \item Filter out the results where the groundtruth samples are assessed with anything except 4 (good) and 5 (excellent), or the samples without voice are assessed with       anything except 1 (bad).
    \item Split randomly the remaining ratings for each model into 5 almost-equal-size groups, compute their mean and std.
\end{enumerate}
Since the models are distributed uniformly among the pages, assessor's biases affect all models in the same way, so the relative order of the models remains.
On the other hand, assessor will have access to all variety of the models on one page and thus can scale his ratings better.
The other side is that the models rating are not independent from each other in this setting, because assessors tend to estimate the sample quality relatively to the average sample of the page, i. e. the more bad models are in comparison -- the bigger MOSes are assigned to the good ones.
4 trapping samples per page is also a reasonable choice, because one cannot just random guess the correct answers for these questions.

The drawback of MOS is that sometimes it requires too much assessors per sample to determine confidently which model is better.
The possible solution is to use a simplified version of comparison category rating, i. e. preference test.
This test compares two models, assessor is asked to chose which model produces the best output for the same input.
If assessor doesn't hear the difference, the option ``equal'' must be selected.
\begin{enumerate}
    \item Select a subset of 40 random samples from the test set.
    \item Randomly shuffle this set split it into the pages of size 20.
    \item Select randomly 10 positions on each page where Model1's prediction will be first.
    \item Insert additional 4 trapping samples into random locations on each page: each trapping sample is a pair of a clean speech from groundtruth and its noticeable distorted version.
    The order of models in trapping sample is random, but on each page there are 2 samples with one order and 2 samples with another.
    \item Upload the pages to the crowd-sourcing platform, set the number of assessors for each page to at least 30.\\
          Assessors are asked to work in headphones in a quiet environment; they must listen the audio until the end before assess it.
    \item Filter out the results where the trapping samples are classified incorrectly.
    \item Use sign test to reject the hypothesis that the models generate the speech of the same [median] perceptual quality.
\end{enumerate}

\begin{figure}[h]
\begin{center}
\includegraphics[width=0.9\linewidth]{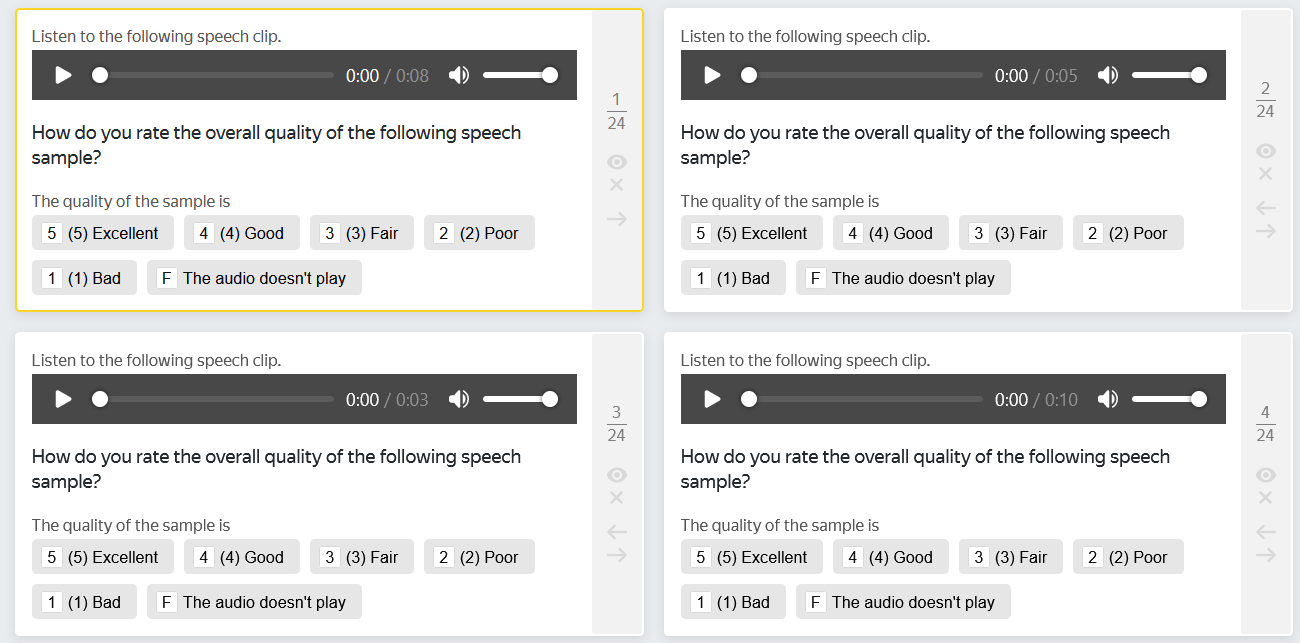}
\end{center}
\caption{The assessor's interface.}
\label{fig:instructions}
\end{figure}

\begin{figure}[h]
\begin{center}
\includegraphics[width=0.9\textwidth]{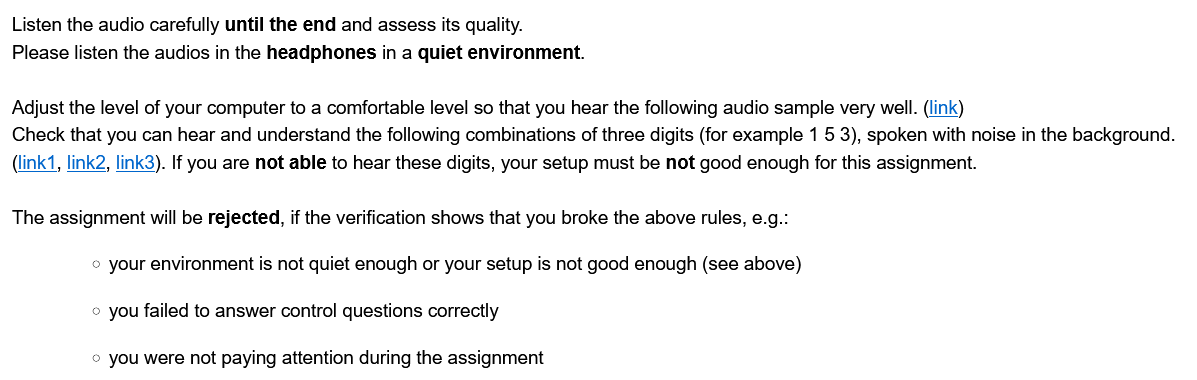}
\end{center}
\caption{The rules for the assessor.}
\label{fig:interface}
\end{figure}

The instructions for the assessors and the screenshots of evaluation interface are provided on \cref{fig:instructions} and \cref{fig:interface} respectively.

The total amount of money spent on the surveys while preparing the paper is somewhere between 1000\$ and 1500\$ (it is hard to estimate exactly because there were a lot of exploration tests during creating the model and writing the paper). According to the crowdsourcing system statistics, the average hourly wage varies between tasks from 2.5\$ to 4\$, which exceeds by large margin the minimal wage in the countries where the test was conducted.





\section{Implementation details}
\label{app:impl}
\subsection{Total amount of compute resources}
We run our experiments mainly on Tesla P40 GPUs. We used approximately 20000 hours for the GPU hours. Note that the significant computational time was consumed to reproduce prior results.

\subsection{Results reproduction}
As a part of this submission supplementary material, we provide all source codes that are needed to train and infer our models. We also attach configuration files that contain all the necessary information regarding the model's specification and hyperparameters.

\subsection{Baselines}

We re-implement the 2S-BWE model \cite{lin2021two} closely following the description provided in the paper. We adopt the method for 1 kHz and 2kHz input bandwidths following the same logic as in 4 kHz. The first minor difference for these cases is that we mirror phase spectrogram 2 times for 2 kHz (from 2 kHz to 4 Khz, then the resulting spectrogram is mirrored again to obtain 8 kHz bandwidth) and analogously 3 times for 1 kHz. The second difference is that the TCN network needs to output more channels for 1 kHz and 2 kHz as the larger spectrogram needs to be predicted. Thus, we change the number of the output channels of the last TCN layer. All 2S-BWE models are trained with the same set of hyperparameters described in the paper.

We implement the SEANet model following the original paper \cite{tagliasacchi2020seanet}. For a fair comparison with our work, we did not restrict the model to be streaming and did not reduce the number of channels as described by \cite{li2021real}. We run the SEANet model for the same number of iterations as the HiFi++, besides that we use the same hyperparameters as described in the original paper.

We also tried to reproduce results of \cite{kim21h_interspeech} (SE-Conformer). Noteworthy, we didn't obtain metrics that are close to the reported in the article.

For comparison with HiFi-GAN \cite{kong2020hifi}, MelGAN \cite{kumar2019melgan}, Waveglow \cite{prenger2019waveglow},
TFilm \cite{birnbaum2019temporal}, VoiceFixer \cite{liu2021voicefixer}, MetricGAN+ \cite{fu2021metricgan+}, DEMUCS \cite{defossez2019music} we employ the official implementations provided by the authors.

\section{Discriminator Analysis}
\begin{figure}[h]
\begin{center}
\includegraphics[width=0.9\linewidth]{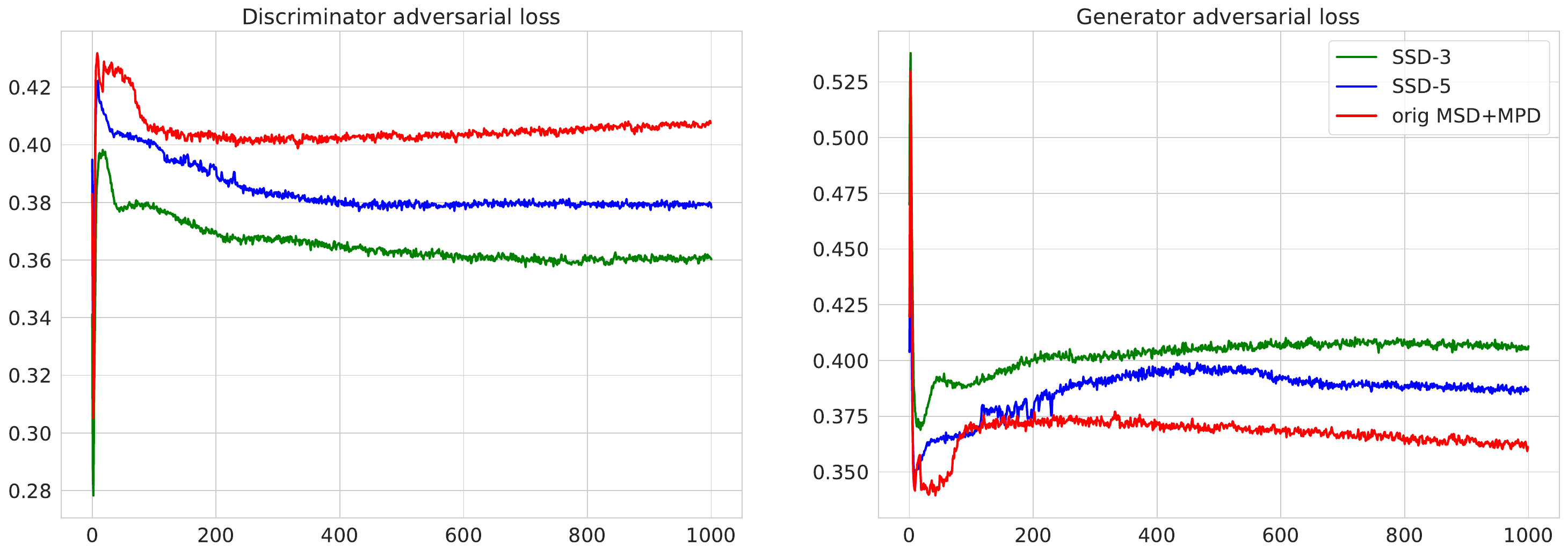}
\end{center}
\caption{Training learning curves for different discriminators normalized to the number of discriminators: SSD-3 (3 SSD discriminators), SSD-5 (5 SSD discriminators), orig MSD+MPD (original 3 MSD and 5 MPD discriminators).}
\label{fig:adv_losses}
\end{figure}

In \Cref{fig:adv_losses} we provide the training learning curves for SSD discriminators and for the original MSD and MPD discriminators (normalized to the number of discriminators). We see that qualitatively the learning process for these settings are quite similar, so we can expect the comparable quality of the trained model that we observe it in the practice.

\end{document}